\documentclass[twocolumn,groupedaddress]{revtex4}
\usepackage{amssymb}
\usepackage{graphicx}

\begin{document}

\title{Depinning and dynamics of AC driven vortex lattices in random media}
\author{D. P\'{e}rez Daroca}
\email{daroca@df.uba.ar}
\affiliation{Departamento de F\'{\i}sica, FCEyN, Universidad de Buenos Aires and IFIBA,
CONICET; Pabellon 1,Ciudad Universitaria, Buenos Aires, Argentina.}
\author{G. S. Lozano}
\affiliation{Departamento de F\'{\i}sica, FCEyN, Universidad de Buenos Aires and IFIBA,
CONICET; Pabellon 1,Ciudad Universitaria, Buenos Aires, Argentina.}
\author{G. Pasquini}
\affiliation{Departamento de F\'{\i}sica, FCEyN, Universidad de Buenos Aires and IFIBA,
CONICET; Pabellon 1,Ciudad Universitaria, Buenos Aires, Argentina.}
\author{V. Bekeris}
\affiliation{Departamento de F\'{\i}sica, FCEyN, Universidad de Buenos Aires and IFIBA,
CONICET; Pabellon 1,Ciudad Universitaria, Buenos Aires, Argentina.}

\date{\today}

\begin{abstract}
We study the different dynamical regimes of a vortex lattice driven by AC forces in the presence of random pinning via numerical simulations. The behaviour of the different observables is charaterized as a function of the applied force amplitude for different frequencies. We discuss the inconveniences of using the mean velocity to identify the depinnig transition and we show that instead, the mean quadratic displacement of the lattice is the relevant magnitude to characterize different AC regimes. We discuss how the results depend on the initial configuration and we identify new hysteretic effects which are absent in the DC driven systems.
\end{abstract}

\pacs{PACS}
\maketitle

\section{\protect\bigskip Introduction}
\label{intro}

A remarkable diversity of  physical systems belong to the category
of driven elastic manifolds moving over random landscapes. Extensively
studied examples are abundant in the literature: moving vortex lattices in
type II superconductors \cite{Jensen, Faleski, Olson2, Kolton, Moon,
Balents, LeDoussal, Kosh-Vino, Olson,Fangohr,Ryu,Bhatta},
 sliding colloidal particles \cite{CRei2,ReichhardtAC} or  charge density wave systems \cite{Balents2}, magnetic bubble
arrays \cite{Seshadri},  driven Wigner crystals \cite%
{Cha,CRei} and  stripe forming systems \cite{CRei3}. 
Among the topics that have recently received much attention is the nature of
the depinning mechanisms that occur as the external DC driving \ force is
increased beyond the critical force $F_{C}^{DC}$, and its relation with
proliferation of topological defects  in the form of bounded or unbounded
disclinations. The dependence of the dynamics with initial conditions has
been examined in detail and it was determined that for DC drives,
memory of initial conditions is lost at depinning \cite{Chandran}. A dynamical depinnig transition has been 
identified at a given force $F_{P}^{DC}$. For larger forces topological defects heal
 and smectic linear flow is observed \cite{Kosh-Vino,LeDoussal}.

 It should be noted however, that most of the research thrust has
been specially focused on DC drives. In experiments, an AC field is often applied to order
the vortex lattice (VL); the most accepted picture is that the AC field
 assists the system in an equilibration process, from a disordered
metastable configuration to the equilibrium Bragg Glass phase, free of
dislocations \cite{Beidenkopf}. However, a large amount of results, in both experiments \cite{Pasquini,Valenzuela2}
 and simulations \cite{Valenzuela, ReichhardtAC,
ReichhardtAC2,Kolton2,Kolton3}, show that, in some cases, an AC drive can
disorder the VL, and that oscillatory dynamics plays an essential role.

The oscillatory dynamics of the VL is in itself a broad field that is not
completely understood. The main porpuse of this work is to provide a more
comprehensive description, using numerical simulations, of AC driven vortex
lattices over a random distribution of pinning sites. We explore the effect
of the applied force amplitude for different
frequencies, starting from different initial configurations. The VL mean
velocity, which can be dephased from  the excitation, is no longer an
adequate observable to detect depinning and we show that instead, the mean
quadratic displacement of the lattice is the relevant magnitude to identify
depinning and dynamical AC regimes. The depinning transition becomes a crossover as is described below.
The pinned linear Campbell regime \cite{Campbell} is also identified and simulations are compared
with analytical calculations. 

The paper is organized as follows: In Sec. \ref{numerical} we describe in
detail the numerical simulations; starting with the model (\ref{model}), following with the numerical
procedures (\ref{procedures}) and giving the definition of the observables (\ref{ob}). In Sec. \ref{results} we present and discuss the results and in Sec. \ref{con} we highlight  the main conclusions.

\section{Numerical simulations}
\label{numerical}

\subsection{The model}
\label{model}

In our simulations, we consider $N_{v}$ rigid vortices with coordinates $%
\mathbf{r}_{i}$ in a two-dimensional rectangle of size $L_{x}\times L_{y}$
that evolve according to the dynamics 

\begin{equation}
\mathbf{F}_{i}-\eta \mathbf{v}_{i}=0
  \label{equ1}
\end{equation}%
where $\mathbf{v}_{i}$ its velocity and $\eta $ the Bardeen-Stephen
viscosity coefficient and $\mathbf{F}_{i}$ is given by the sum of the
vortex-vortex interaction, the pinning attraction and the Lorentz force 

\begin{equation}
\mathbf{F}_{i}=\mathbf{F}_{i}^{vv}+\mathbf{F}_{i}^{vp}+\mathbf{F}_{i}^{L}.  
\label{equ2}
\end{equation}%

The vortex-vortex interaction per unit length is given by, 

\begin{equation}
\mathbf{F}_{i}^{vv}=\sum_{j\neq i}^{N_{v}}\mathbf{F}^{vv}(\mathbf{r}_{i}-%
\mathbf{r}_{j}),
\end{equation}%
where 

\begin{equation}
\mathbf{F}^{vv}(\mathbf{r}_{i}-\mathbf{r}_{j})=\frac{\phi _{0}^{2}}{8\pi^2 
\lambda ^{3}}f_{vv}K_{1}(\frac{\mid \mathbf{r}_{i}-\mathbf{r}_{j}\mid }{%
\lambda })\hat{\mathbf{r}}_{ij}.
\end{equation}%
Here, $\phi _{0}$ is the quantum of magnetic flux, $\lambda $ is the London
penetration length and $K_{1}$ the
special Bessel function. The parameter $f_{vv}$ is dimensionless and can be
related to the stiffness of the vortex lattice (see Ref. \onlinecite{Valenzuela}). The 
$N_{p}$ pinning centers are supposed to be located at random positions $%
\mathbf{R}_{j} $, and their interaction with vortices is modeled by 

\begin{equation}
\mathbf{F}^{vp}(\mathbf{r}_{i})=\sum_{j=1}^{N_{p}}=\mathbf{F}^{vp}(\mathbf{r}%
_{i}-\mathbf{R}_{j}),
\end{equation}

\begin{equation}
\mathbf{F}^{vp}(\mathbf{r}_{i}-\mathbf{R}_{j})=-F_{j}^{p}e^{-(\frac{%
|r_{i}-R_{j}|}{r_{p}})^{2}}({\mathbf{r}}_{i}-{\mathbf{R}}_{j}),
\end{equation}%
here, $F_{j}^{p}$ (chosen from a Gaussian distribution with mean value $F^p = 0.2$ and a standard deviation of $0.1F^p$) and $r_{p}$ tune
the strength and range of the interaction.

The Lorentz force per unit length is given by $F_{L}=\phi _{0}J_{ext}\times
z $ where $J_{ext}$ is the external driving current density and $z$ is the
versor perpendicular to the plane.

Following Ref. \onlinecite{Valenzuela} we measure lengths in units of $\lambda $,
forces (per unit length) in units of $f_{0}=\frac{\phi _{0}^{2}}{8\pi^2
\lambda ^{3}}$, time in units of $t_{0}=\eta \lambda /f_{0}$ and frequency in units of $\omega_0 = 1/t_0$. In our
simulation, we will consider $N_{v}=1600$, $L_{x}=40\lambda $, $L_{y}=\sqrt{3%
}L_{x}/2$, $N_{p}=25N_{v}$ and $r_{p}=0.2\lambda $.

Concerning the numerical details, the equations of motion are
 integrated using a standard Euler algorithm with step $h=0.04t_{0}$, and
a hard cut-off $\Lambda =4\lambda $ in calculating the vortex-vortex force.
When calculating physical observables we average over four realizations of disorder.

\subsection{Procedures}
\label{procedures}

The AC drive is simulated with an external square AC force of the form 
\begin{equation}
F_{L}^{AC}=\left\{ 
\begin{array}{ccc}
f_{L} & if & nT<t<nT+\frac{T}{2} \\ 
-f_{L} & if & nT+\frac{T}{2}<t<(n+1)T%
\end{array}%
\right.
\end{equation}

We study vortex lattice evolution as we vary the force in a slow ramping of $%
f_{L}$ starting from different initial conditions. In each case, we leave
the system evolving\ from the initial configuration with zero external force
for 5 cycles. This free evolution creates or annihilates a low
density of defects and becomes our metastable initial state. A numerical
realization will be obtained by ramping from $f_{L}$ to $f_{L}+\Delta f_{L}$%
, applying $N_{a}$ cycles this new force, and then allowing the
system to relax for $N_{w}$ cycles at zero force, before reassuming 
 the ramping. Measuring is performed during the last cycle, 
just before switching off the applied force (i.e. during cycle $N_{a}$).

\subsection{Observables}
\label{ob}

It is by now well known that the mobility of the VL is affected by
the topology of the VL configuration . A common observable used to
characterize the VL configuration is the density of lattice defects, $n_{d}$, (i.e. vortices with 5 or 7 neighbors in the Delaunay triangulation).

The mobility itself is characterized by the vortex velocity. The
instantaneous mean vortex velocity in the direction of the applied force at
time $t$ is 

\begin{equation}
v(t)=\frac{1}{N_{v}}\sum_{i}^{N_{v}}v_{i}(t).
\end{equation}

In the DC case, the $v$ vs $F_{L}$ curves can be directly
related with experimental current-voltage characteristics ($V-I$ curves).

In the AC case the relationship between velocity and force
and the experimental $V-I$ curve is more subtle, as the phase
between the applied force and the velocity plays an essential role.

We then define the half cycle mean velocity in the $n^{th}$ cycle as

\begin{equation}
v_{n}=\frac{2}{T}\int_{nT}^{nT+T/2}v(t)dt  \label{vn}
\end{equation}

Notice that the half cycle velocity takes into acount this phase
factor. We estimate the mean phase factor $\varphi _{n}$ by applying a
sinusoidal $F_{L}(t)$ and calculating the phase between $F_{L}$ and the
first harmonic of the mean velocity $v(t)$ in the $n^{th}$ cycle.

Moreover, a quantity of interest is the average quadratic displacement
defined as 

\begin{equation}
<\delta X^{2}>(t_{2},t_{1})=\frac{1}{N_{v}}%
\sum_{i=1}^{N_{v}}(x_{i}(t_{2})-x_{i}(t_{1}))^{2}. 
 \label{x2}
\end{equation}%

In the AC case, we can take $t_{2}=t_{1}+pT$, the mean quadratic displacement in the
force direction after performing $p$ oscillations. This quantity, as we will show below,
is very useful for describing AC drives.

\section{Results and discussion}
\label{results}

Throughout the work we have explored the behaviour of the system
as a function of the AC amplitude. Recent numerical studies on a similar
system (colloids under the influence of an external AC drive and quenched
disorder) \cite{ReichhardtAC} have found a crossover from a low drive-highly
disorderd phase to a high drive-low disordered phase. In all cases
considered in Ref. \onlinecite{ReichhardtAC} the initial configuration was
 highly disordered. Our results confirm the existence of the
high drive-low disorderd phase but we will show that when more general
initial conditions are considered then richer, low drive phases emerge.

\begin{figure}[h]
\includegraphics[height=100mm]{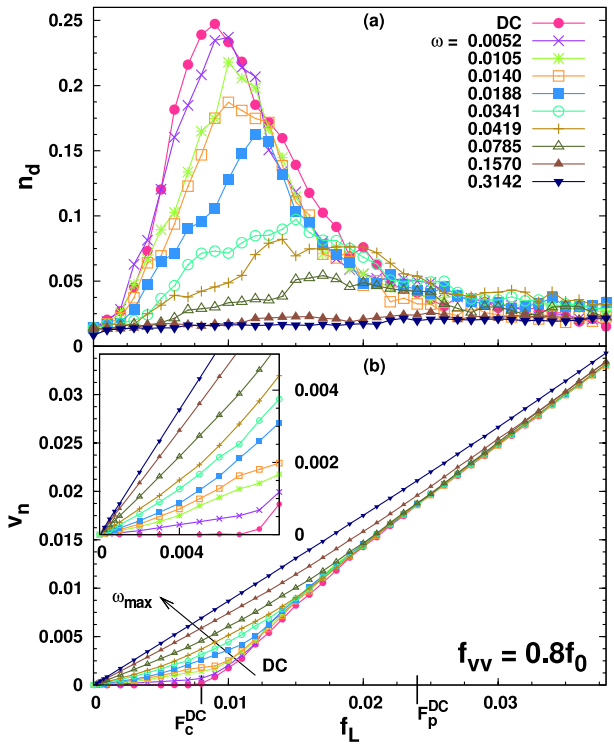}
\caption{ {\protect\footnotesize Number of topological defects $n_{d}$ (a) and half cycle mean velocity $v_{n}$ (b) (see text) as a function of the amplitude of the applied Lorentz force $f_{L}$ for different frequencies $\protect\omega $. All the curves are calculated slowly ramping $f_{L}$, from an ordered VL configuration. Inset: zoom of $v_{n}(f_{L})$ curves at low amplitudes: $v_{n}>0$ at finite frequency.}}
\label{ndvsFL_frec}
\end{figure}

\begin{figure}[h]
\includegraphics[height=100mm]{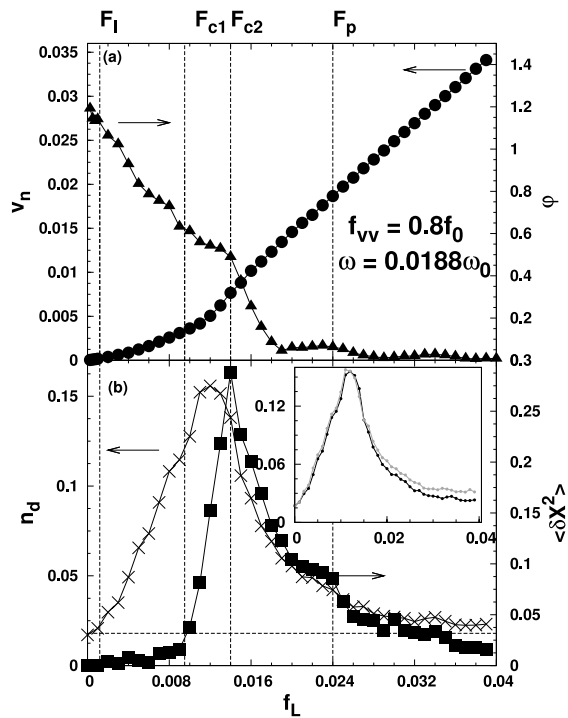}
\caption{ {\protect\footnotesize $v_{n}$ (a-left),  phase factor $
\protect\varphi $ (a-right), $n_{d}$ (b-left) and mean quadratic
displacement $<\protect\delta X^{2}>$ (b-right) as a function of $f_{L}$ at $
\protect\omega =0.0188\omega_0$. Vertical dashed lines indicate the different AC
regions discussed in the text. The depinning of the VL occurs in a crossover
region between $F_{c1}$ and $F_{c2}$, where $<\protect\delta X^{2}>$
grows. Inset: Comparison between $n_{d}$ before (black full circles) and after
(grey full circles) $N_w$ cycles without applied force (see text). Beyond the
depinning region the moving lattice \ configurations are unstable and relax. }}
\label{multiT333}
\end{figure}

Results starting from an ordered configuration are summarized in Figure \ref{ndvsFL_frec}
and Figure \ref{multiT333}. In all the cases $N_{a}=6$ cycles have been applied with
each $f_{L}$ force. The observables $v_{n}$ (Eq. \ref{vn}) and $n_{d}$ have
been calculated in the $n^{th}$cycle before switching off $f_{L}$. \ In
Figure \ref{ndvsFL_frec} we show the behaviour of the defect density $n_{d}$ (Fig. \ref{ndvsFL_frec}(a)) and the
characteristic $v_{n}-f_{L}$ curve (Fig. \ref{ndvsFL_frec}(b)) for a particular choice of
vortex-vortex interaction ($f_{vv}=0.8f_0$), at various frequencies $\omega $.
In the DC limit (shown in red full circles) we clearly identify three
different regimes, a pinned lattice for $f_L<F_{c}^{DC}$, a disordered
flow (plastic) region for $F_{c}^{DC}<f_{L}<F_{p}^{DC}$ and a flowing
(smectic) linear regime for $f_{L}>F_{p}^{DC}$ consistent with previous 
work \cite{Chandran}. While the existence of $F_{p}^{AC}$ can be inferred form these AC
curves, as the value for which all curves that present plastic motion merge,
the distinction between the pinned and plastic regimes is less apparent.
This is so, because even a weak driving force produces oscillations of
vortices around their pinning centers with finite mean velocity. This is a strong argument to abandon the mean velocity as
the adequate observable to indicate depinning. The inset of Fig. \ref{ndvsFL_frec}b
shows an enlarged area for small driving forces: at very small AC amplitudes
a linear Campbell regime holds but the response becomes non-linear at
amplitudes well bellow $F_{c}^{DC}$. The definition of a critical force in
the AC response is therefore not so obvious, and we will discuss this point
in detail later.

In the AC curves, we distinguish two frequency regimes: at low frequencies
($\omega \lesssim 0.02\omega_0$ in our simulation), restoring and pinning forces
 prevail over losses, whereas at high frequencies viscous
drag dominates. The low frequency regime is characterized by a highly
nonlinear response at intermediate AC amplitudes, associated with a
pronounced peak in $n_{d}$. This feature smears at higher frequencies, the
density of topological defects decreases, and the non-linearity is less
apparent, going to a linear \textquotedblleft ohmic\textquotedblright\
regime (blue down triangles) at very high frequency ($\omega \gtrsim 0.2\omega_0$)
where dissipative forces govern. Memory and history effects occur when
pinning and elastic forces compete and both prevail over viscous forces.
Therefore, in the following analysis we will focus in the low frequency
regime.

In Figure \ref{multiT333} we plot together various observables as a function of the AC
amplitude for a fixed frequency $\omega =0.0188\omega_0$, in a procedure analogous to
that explained in \ the description of Figure \ref{ndvsFL_frec}. Figure \ref{multiT333}(a) displays the
modulus of the half cycle mean velocity $v_{n}$ (left axis) and the
estimated phase factor $\varphi _{n}$ (right axis). In Figure \ref{multiT333}(b) the density
of defects $n_{d}$ (left axis) and the mean quadratic displacement $<\delta X^{2}>$  after 5 cycles (right axis) are shown. The dashed horizontal line in
Figure \ref{multiT333}(b) indicates de density of defects $n_{dr}$ that are spontaneously
created (without any applied force), and correspond to the initial more
ordered configuration. A still more ordered configuration is unstable.

We identify various regimes. At very low AC amplitudes ($f_{L}<F_{l}$) a
linear $v_n(f_{L})$ holds. The velocity is mainly out of phase. The VL
configuration remains unchanged, because vortices are trapped and can only
perform small harmonic oscillations around their initial positions. This
motion is reversible, and therefore $<\delta X^{2}>=0$. In the other
limit, at very high AC amplitudes ($f_{L}>F_{p}$), there is a dynamic
reordering and the pinning potential is completely smeared. There is a
linear in phase Ohmic response.

In the intermediate range $F_{l}<f_{L}<F_{p}$, the response is
non-linear. A very rich behaviour with different non-linear regimes may be
observed. In the first non-linear region, at small amplitudes ( $%
F_{l}<f_{L}<F_{c1}$), $\ $the velocity is small, and there is not an
appreciable displacement in the direction of the force (i.e. $<\delta X^{2}>\sim 0$); plastic random displacement produces a huge number of
dislocations. However, most of the vortices remain trapped around
the pinning sites. Depinning occurs in a crossover region, between $F_{c1}$ and $F_{c2}$. We identify the beginnig of the depinning region at the force $F_{c1}$, where vortices move in average distances larger than the pinning radius $r_{p}$. In this region, the irreversible displacement grows with $f_{L}$ and is reflected in the growth of $<\delta X^2>$, that reaches its maximun at the second force $F_{c2}$; as this happens, the density of defects attains its maximum and  the slope of $v_{n}(f_{L})$ grows.  Above $F_{c2}$, in the upper non-linear region $(F_{c2}<f_{L}<F_{p})$, the mean phase factor drastically decreases (i.e. dissipation becomes relevant), while the motion becomes  more reversible and the VL more ordered. This is accompanied by a smooth decrease in the $v_{n}(f_{L})$ slope that approaches the final linear relationship. 
Another interesting point to remark is that VL configurations that have similar
density of defects at both sides of the depinning transition are
qualitatively different. This difference can be detected, observing the
relaxation of the VL configurations after removing the applied force. In the
inset of Figure \ref{multiT333}(b), we compare the density of defects $n_{d}$ before switching
off the applied force (black full circles) with the same observable, $N_{w}$ cycles after removing the applied force (grey full circles).
While for forces lower than $F_{c2}$, the vortices move around pinning sites
in robust metastable configurations, beyond the depinning transition the
moving lattice configurations are unstable, and relax towards a state with more defects.

\begin{figure}[h]
\includegraphics[height=100mm]{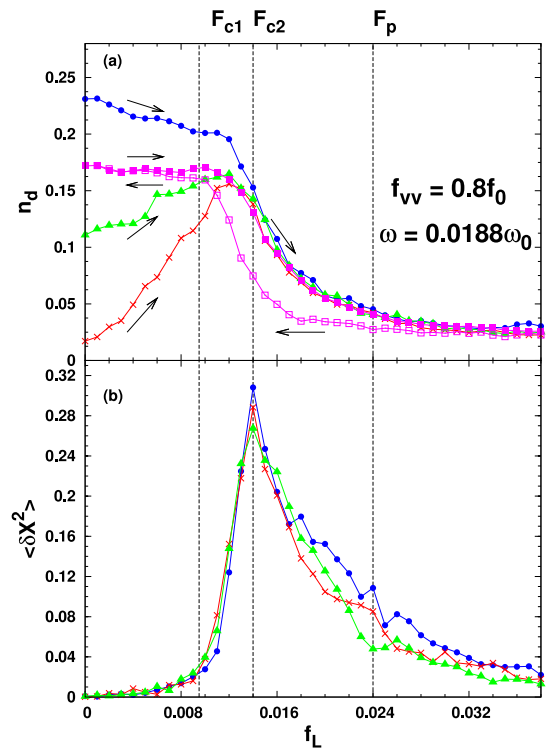}
\caption{ {\protect\footnotesize (a) $n_{d}$ as a function of $f_{L}
$ starting from different initial configurations. Arrows indicate the 
direction of change of $f_{L}$. Above $F_{c2}$ all the $n_{d}(f_{L})$ curves
taken with increasing $f_{L}$ merge. An hysteresis is observed (full and empty magenta squares).
 (b) $<\protect\delta X^{2}>(f_{L})$ curves corresponding to the
same processes described in (a). This observable is independent of the
initial condition for the whole driving force range. }}
\label{multiDifIni}
\end{figure}

We now turn to analyze the various AC regimes starting from different
initial conditions. In Figure \ref{multiDifIni} the density of defects $n_{d}$ (Fig. \ref{multiDifIni}(a))  and
the mean quadratic displacement $<\delta X^{2}>$ (Fig. \ref{multiDifIni}(b)) as a function of the AC
force are shown for different initial conditions and AC histories. The
results described in the previous paragraph (starting from an ordered VL and
increasing the AC force) are plotted in red crosses. Analogous
results obtained with a similar AC protocol but starting from the more
disordered metastable configuration are plotted in blue dots, and results
with an initial intermediate $n_{d}$ are plotted in green triangles. As
in the DC case studied in Ref.  \onlinecite{Chandran}, the different $n_{d}$ curves
merge for $f_{L}$ larger than a threshold value. For the DC case, this
threshold value was identified with $I_{c}^{DC}$, the critical current at
depinning. As we mentioned before, the identification of the critical
current $I_{c}^{AC}$ is less obvious in the AC case, and we notice that
the $n_{d}(f_{L})$ curves merge around $F_{c2}$. In magenta squares, we show
the behaviour of $n_{d}$ for decreasing (open squares)
and a subsequent increasing (full squares) external driving. We observe the
presence of hysteresis in the region between $F_{c1}$ and $F_{p}$ . Below $%
F_{c1}$ pinning forces dominate, and above $F_{p}$ they are completely
smeared. In the intermediate region, the strong competition gives rise to
hysteresis. This hysteretic effect beyond the critical force has not been
observed in the DC case\cite{Chandran} and it seems to be a new geninue effect of the AC case.

As another striking result, the behaviour of $<\delta X^{2}>$ vs. $F_{L}$
is independent of the initial condition for the whole driving force range.
The reorganization of vortex defects by AC drives in initially different
configurations, and for different AC protocols, surprisingly does not
involve different quadratic displacements. This observable is
univocally related with the applied force, and seems to be the best one to
characterize the AC regimes.

\begin{figure}[h]
\includegraphics[width=80mm] {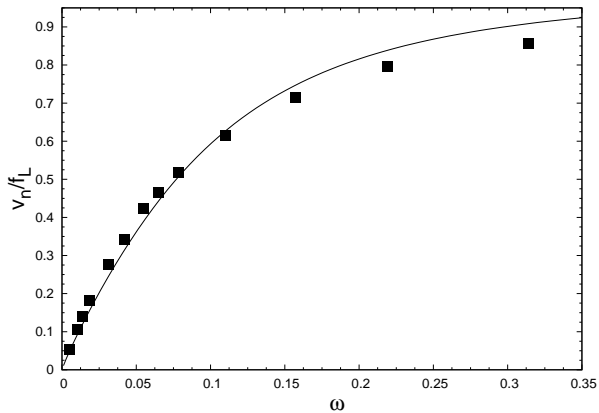}
\caption{ {\protect\footnotesize Normalized half cycle mean
velocity $v_{n}/f_{L}$ as a function of frequency $\protect\omega $ in the
linear regime, at low $F_{L}.$  Fit with Eq.\protect\ref{eqfit}  (continuos
line), considering a Gaussian distribution for the Labusch constant,  gives $%
\frac{\protect\alpha _{L}}{\protect\eta }\sim \protect\omega _{c}=0.11\pm
0.05$.}}
\label{fit}
\end{figure}

Finally, we discuss in more detail the linear regime obtained at very small
AC driving forces. In the harmonic approximation we can write the mean
restoring force (averaging $\mathbf{F}_{i}^{vv}+\mathbf{F}_{i}^{vp}$ in Eq. \ref{equ2}) as $\mathbf{F}_{rest}\mathbf{=-\alpha {x}}$ where $\mathbf{x}$
denotes the average displacement of vortices from their equilibrium position
and $\alpha $ is the Labusch parameter \cite{Campbell}. Within this aproximation, Eq. \ref{equ1}, can be written as: 
\begin{equation}
\eta \dot{x}+\alpha x=F_{L}^{AC}.
  \label{aprox}
\end{equation}

If we consider a semicycle for which $F_L^{AC}$ is positive, the solution of
equation (\ref{aprox}) is 
\begin{equation}
x(t) = \frac{f_L}{\alpha}+ (x_0(t_n)-\frac{f_L}{\alpha}) e^{-\frac{\alpha}{%
\eta}(t-t_n)},
\end{equation}
where $nT<t<nt+\frac{T}{2}$.

This equation provides a recurrence relation for $x_0(t_n)$ that can be
solved explcitly. In the limit $n \longrightarrow \infty$ (the stationary
regime) this gives 
\begin{equation}
x_0(t_n) = -\frac{f_L}{\alpha} \tanh(\frac{\alpha T}{4 \eta})
\end{equation}
and the average velocity over half-cycle as 
\begin{equation}
\overline{v_n} = \frac{2\omega f_L }{\alpha \pi} \tanh(\frac{\alpha \pi}{2
\eta \omega}).
  \label{eqfit}
\end{equation}

On the other hand, numercially we have performed different calculatons of
the half cycle mean velocity in the $n^{th}$ cycle $v_{n}$ (according to Eq. \ref{vn}) as a function of $\omega $ for different values of $f_{L}$, so we can
average over $f_{L}$ and study the behaviour of $\overline{v_n}/f_{L}$. In
addition, as we did before, we have performed averages over different
realizations of pinning centers. This fact implies that indeed $\alpha $ is a
random variable that as a first aproximation can be assumed to be Gaussian
distributed and characterized by its mean value $\overline{\alpha }$ and
standard deviation $\sigma _{\alpha }$.

In Figure \ref{fit} we show the results of the simulation, where we have used 7 values of $f_{L}$ from $0.00025f_0$ to $0.004f_0$ and 4 realizations over disorder. Our fit gives $\overline{\alpha} =0.11,$ $\sigma_{\alpha} =0.05$. \ 

 A critical frequency may be estimated as $\omega _{c}\sim \frac{\overline{\alpha} }{\eta }=0.11\pm 0.05$. We are able now to justify the two frequency regimes introduced at the beginning of this section.

 \section{Conclusions}
\label{con}

We have presented a comprehensive description of the dynamics of AC driven
vortex lattices over a random distribution of pinning sites, focusing the
study in the low frequency regime, $\omega <<\omega _{c}$, where pinning and
elastic forces prevail over viscous forces. This critical frequency $\omega
_{c}$ has been estimated by fitting results of numerical simulations at low
AC amplitudes with that predicted by an analytical model.

The reorganization of vortex defects by AC drives from initially different
configurations and for different AC protocols produce history dependent
vortex lattice configurations with different density of defects and mobility.

As an important result, we notice that the behaviour of the mean quadratic
displacements $<\delta X^{2}>(f_{L})$ is independent of the initial
condition for the whole driving force range. This observable is univocally
related with the applied force, and seems to be the best one to characterize
the AC regimes. This should be contrasted with the mobility that can be large even in pinned AC driven lattices.

As expected, a linear Campbell regime holds at very low AC amplitudes ( $%
f_{L}<F_{l}$\ \ ) whereas the response is Ohmic at very high AC amplitudes ($%
f_{L}>F_{p}$ \ ) where there is a dynamic reordering and the pinning
potential is completely smeared. In both linear responses the motion is
reversible and $<\delta X^{2}>=0.$

In all the intermediate range $F_{l}<f_{L}<F_{p}$, the response is
non-linear and a very rich behaviour with different non-linear regimes may be
observed.

Depinning occurs in a crossover region ($F_{c1}<f_{L}<F_{c2}$) where $%
<\delta X^{2}>$ grows with $f_{L}$ and vortices move in average distances
much larger than the pinning radius.

In the first non-linear region, below the depinning transition ( $%
F_{l}<f_{L}<F_{c1}$) there is not an appreciable displacement in the
direction of the force (i.e. $<\delta X^{2}>\sim 0$); plastic random
displacement produces a huge number of dislocations but most of the
vortices remain trapped around the pinning sites. The density of defects $%
n_{d}$ is strongly dependent on the initial configuration.

Beyond the depinning region the memory of the initial configuration is
lost. Dynamics becomes more and more reversible as $<\delta X^{2}>$
decreases. Dissipation becomes relevant and the VL reorders. The moving
vortex lattice configurations become unstable and relax after switching off
the applied force. However, there is a large region ( $F_{c2}<f_{L}<F_{p}$)
where plastic motions are still present, as can be inferred from $\
<\delta X^{2}>>0$ and the non-linear response. Below the plastic threshold $%
F_{p}$, an hysteresis in the applied force, absent in DC driven lattices, is
observed, showing once again the rich particularities of oscillatory dynamics.

\begin{acknowledgments}
This work was partially supported by CONICET PIP 1212/09 and PIP 112 200801 00930, UBACyT X13, X123 and X166.
\end{acknowledgments}

\end{document}